\documentclass[prd,aps,preprintnumbers,nofootinbib,twocolumn]{revtex4}
\usepackage{epsfig}
\usepackage{amsmath}
\usepackage{hyperref}
\usepackage{physics}
\usepackage{xcolor}
\usepackage{accents}

\begin{document}

\preprint{P3H-22-121, TTK-22-44}

\renewcommand{\thefigure}{\arabic{figure}}

\title{ A study of additional jet
  activity in top quark pair production \\ and decay at the LHC}

\author{Giuseppe Bevilacqua}
\affiliation{{\small  Institute of Nuclear and Particle Physics, NCSR
    "Demokritos", 15341 Agia Paraskevi, Greece}}
\author{Michele Lupattelli}
\affiliation{{\small Institute for Theoretical Particle Physics and
    Cosmology, RWTH Aachen University, D-52056 Aachen, Germany}}

\author{Daniel Stremmer}
\affiliation{{\small Institute for Theoretical Particle Physics and
    Cosmology, RWTH Aachen University, D-52056 Aachen, Germany}}

\author{Ma\l{}gorzata Worek}
\affiliation{{\small Institute for Theoretical Particle Physics and
    Cosmology, RWTH Aachen University, D-52056 Aachen, Germany}}

\date{\today}

\begin{abstract}
We report on the calculation of the NLO QCD corrections to top quark
pair production in association with two hard jets at the LHC.  We take
into account higher-order effects in both the production and decays of
the top-quark pair. The latter are treated in the narrow width
approximation and $t\bar{t}$ spin correlations are preserved
throughout the calculation.  This is the first time that such a
complete study for this process is conducted at NLO in QCD. We present
results for fiducial cross sections at the integrated and differential
level. Furthermore, we investigate kinematic properties of the
additional light jets and their distribution in the $pp\to t\bar{t}jj$
process. We examine jets in production and top-quark decays as well as
the mixed contribution, where the two hardest non-$b$ jets are present
simultaneously in the production and decay processes.

\end{abstract}

\maketitle

High precision theoretical predictions for $t\bar{t}$ production are
crucial for precise measurements of the cross section and top-quark
properties, which are carried out at the Large Hadron Collider (LHC)
\cite{CMS:2018adi,CMS:2019esx,CMS:2019nrx,ATLAS:2019hau}. About half
of the $t\bar{t}$ events, however, are accompanied by additional hard
jet(s) arising from QCD radiation. Such events contribute to truly
multiparticle final states, which are currently measured at the LHC
\cite{ATLAS:2016qjg,ATLAS:2018acq,CMS:2020grm}.  Good understanding
and excellent theoretical control of extra jet activity is a key for
the entire top-quark physics program.  In addition to its importance
as a signal process, it turns out that the $t\bar{t}$ plus jets
production can also be an important background process. For example,
it is essential in Standard Model (SM) measurements of associated
Higgs-boson production with a $t\bar{t}$ pair, where the Higgs boson
decays into a $b\bar{b}$ pair. In addition, multi-particle final
states originating from $t\bar{t}jj$ are abundantly predicted in
various supersymmetric theories. Finally, anomalous production of
additional jets accompanying a $t\bar{t}$ pair could be a sign of new
physics. In all these scenarios, a small signal must be extracted from
the overwhelming SM $t\bar{t}jj$ background.

For processes involving particles interacting via the strong
interaction at least next-to-leading order (NLO) QCD predictions are
needed to describe them in a reliable way. The NLO QCD corrections to
$pp\to t\bar{t}jj$ production with stable top quarks have been
calculated in Refs.  \cite{Bevilacqua:2010ve,Bevilacqua:2011aa}. In
Ref.  \cite{Hoeche:2014qda} $t\bar{t}$ pairs with up to two jets
computed at NLO QCD have been consistently merged with a parton
shower.  In Ref.  \cite{Hoche:2016elu} theoretical predictions for the
production of $t\bar{t}$ pairs with up to even three jets at NLO QCD
were presented.  Finally, in Ref. \cite{Gutschow:2018tuk} the dominant
NLO EW corrections have been incorporated into a parton shower
framework for $t\bar{t}$ plus multi-jet production.  In all of these
studies, however, either only stable top quarks were considered or
top-quark decays were treated in the parton shower approximation
neglecting $t\bar{t}$ spin correlations either at LO or at NLO. The
first case can provide information on the size of the higher-order
effects in production rates but lacks a reliable description of
fiducial-level cross sections. In the second case the parton shower,
which incorporates the dominant soft-collinear logarithmic
corrections, can approximate radiative effects in top-quark
decays. The latter, however, are described by matrix elements formally
accurate at LO only. Incorporating higher-order corrections in the
decay processes already at the matrix element level would therefore be
a further step towards a more complete description of the jet
radiation pattern at NLO.

In this paper, we extend the previous studies for $pp\to t\bar{t}jj$
by adding higher order effects also to top-quark decays to match the
level of accuracy already present in the production stage.
Specifically, we calculate NLO QCD corrections to $t\bar{t}jj$
production and top-quark decays including $t\bar{t}$ spin correlations
utilising the narrow-width approximation (NWA).  This is the first
time that such a complete study for this process is conducted at NLO
QCD. The number of similar calculations for a $2\to 6$ processes (the
decay products of the $W$'s are not counted, because they do not
couple to colour charged states) at NLO in the NWA is rather
limited. The $pp \to t\bar{t}b\bar{b}$ process is the only other case
for which a full NLO description beyond the stable top-quark picture
\cite{Bredenstein:2009aj,Bevilacqua:2009zn,Bredenstein:2010rs,Buccioni:2019plc}
is available \cite{Bevilacqua:2022twl}. Even NLO QCD predictions
beyond the NWA have been calculated in Refs.
\cite{Denner:2020orv,Bevilacqua:2021cit}.  For a less complicated
$pp\to t\bar{t}j$ production the theoretical landscape is much more
satisfactory as both the full off-shell and NWA descriptions are
available \cite{Melnikov:2011qx,Bevilacqua:2015qha,Bevilacqua:2016jfk,
Bevilacqua:2017ipv}.  Furthermore, NLO fixed-order $pp\to
t\bar{t}b\bar{b}$ and $pp\to t\bar{t}j$ calculations have been also
matched to parton shower programs
\cite{Kardos:2011qa,Alioli:2011as,Kardos:2013vxa,
Cascioli:2013era,Garzelli:2014aba,Czakon:2015cla,Bevilacqua:2017cru,Jezo:2018yaf},
and top quark decays (in some cases spin-correlated) have been
included at LO accuracy. It would be beneficial to have a similar
range of predictions for $pp \to t\bar{t}jj$ so that they all can
ultimately be compared to understand their individual strengths and
weaknesses.  The first step in this direction is undertaken in this
work.

In our studies we concentrate on $pp\to t\bar{t}(jj) \to W^+W^- b\bar{b}
jj \to \ell^+ \ell^- \, \nu_\ell \bar{\nu}_\ell \,b\bar{b} \,jj$ decay
chain, where $\ell ^\pm = \mu^\pm, e^\pm$.  We shall refer to the
process as $t\bar{t}jj$, implicitly assuming that decays of the
top-quarks and $W$ bosons are taken into account. By employing
the di-lepton decay channel we can distinguish the additional light
jets from the hadronic top-quark decays. Having included jet emissions
from various stages, we can explore their distribution and impact on
the integrated and differential fiducial cross sections. To be as
realistic as possible, we closely follow the event selection from the
experimental analysis that has recently been carried out by the CMS
collaboration \cite{CMS:2022uae}.

In our calculation top quarks and $W^\pm$ gauge bosons are treated in
the NWA. This approximation is obtained from the full cross section
with unstable particles by taking the limit $\Gamma /m\to
0$. In this way all terms that are less singular than $\Gamma^{-2}$
are consistently neglected, see e.g Refs. \cite{Melnikov:2009dn,
Melnikov:2011qx,Campbell:2012uf,Behring:2019iiv,Czakon:2020qbd}.
Following closely the notation from Ref. \cite{Melnikov:2011qx} we
write the formula for $\sigma^{\rm LO}_{t\bar{t}jj}$ in such a way
that  one can clearly show the different contributions:
\begin{equation}
\label{eq_LO}
\begin{split}
    d\sigma^{\textrm{LO}}_{t\bar{t}jj}&=\Gamma_{t}^{-2}
    \bigl(
    \overbrace{d\sigma_{t\bar{t}jj}^{\textrm{LO}}\,
    d\Gamma_{t\bar{t}}^{\textrm{LO}}}^{\rm Prod.}
    +\overbrace{d\sigma_{t\bar{t}}^{\textrm{LO}}
    \,d\Gamma_{t\bar{t}jj}^{\textrm{LO}} }^{\rm Decay}   
    +\overbrace{d\sigma_{t\bar{t}j}^{\textrm{LO}}
    \,d\Gamma_{t\bar{t}j}^{\textrm{LO}}}^{\rm Mix}
    \bigl)
    \end{split}
\end{equation}
The first term in Eq.  \eqref{eq_LO} describes $t\bar{t}jj$ production
followed by top-quark decays. In this case light jets can occur only
in the production stage. The middle term represents $t\bar{t}$
production followed by top-quark decays with two light jets appearing
only in  decays. The last term is responsible for the mixed
contribution, where light jets appear simultaneously in the
production and decay stage.  We will refer to these contributions as
{\it Prod.}, {\it Decay} and {\it Mix} respectively.

A generalisation to the $\sigma^{\rm NLO}_{t\bar{t}jj}$ case is given
in Eq. \eqref{eq_NLO}, where $\alpha_s$ corrections to each term {\it
Prod.}, {\it Decay} and {\it Mix} have been added.  The symbol "virt"
corresponds to virtual corrections and "real" denotes the case where
one parton is allowed to be unresolved. 

All theoretical predictions have been calculated with the
\textsc{Helac-Nlo} Monte Carlo (MC) program \cite{Bevilacqua:2011xh}.
Specifically, we compute the virtual corrections using
\textsc{Helac-1Loop}
\cite{Ossola:2006us,Ossola:2007ax,vanHameren:2009dr,Draggiotis:2009yb,
vanHameren:2010cp}, which has been extended to provide factorisable
one-loop contributions for $pp\to t\bar{t} +X$ processes
\cite{Bevilacqua:2019quz}. We cross check our results with
\textsc{Recola} \cite{Actis:2012qn,Actis:2016mpe,Denner:2016kdg},
which is a computer program for the automated generation and numerical
computation of tree and one-loop amplitudes.  Because \textsc{Recola}
is also able to provide 1-loop matrix elements in the so-called
double-pole approximation, see e.g.
\cite{Denner:2000bj,Accomando:2004de,Denner:2016jyo}, it was
straightforward to interface it to \textsc{Helac-NLO} and use it in
our case. Furthermore, we have implemented in \textsc{Recola} the
random polarisation method, which replaces the polarisation state with
a linear combination of helicity eigenstates
\cite{Draggiotis:1998gr,Draggiotis:2002hm,Bevilacqua:2013iha}.  This
results in a drastic speed improvement, since now only one (random)
polarisation configuration must be calculated for each event. The
singularities from soft or collinear parton emissions have been
isolated via subtraction methods for NLO QCD calculations. We employ
two independent subtraction schemes, the Catani-Seymour
\cite{Catani:1996vz,Catani:2002hc} and Nagy-Soper scheme
\cite{Bevilacqua:2013iha}, that are both implemented in
\textsc{Helac-Dipoles} \cite{Czakon:2009ss}. In Refs.
\cite{Basso:2015gca,Bevilacqua:2019quz,Bevilacqua:2022twl} the
Catani-Seymour subtraction scheme has been extended to the NWA case to
take into account gluon radiation in the decay of on-shell top quarks.
In order to compute NLO QCD corrections to $pp \to t\bar{t}j(j)$ in
the NWA, however, additional dipoles have been added since the
additional processes $t \to bW^+ gg$ and $t \to bW^+ q\bar{q}$ are
present. We have incorporated the final-initial dipoles for the
splittings $g \to gg$ \cite{Melnikov:2011qx} and $g \to q\bar{q}$ with
the top quark being the spectator. Additionally, the Nagy-Soper
subtraction scheme has been adapted to deal with radiative corrections
in the decay of on-shell top quarks. Up until now this subtraction
scheme was not suitable for calculations in the NWA, thus, this is the
first time we use it for this purpose. The Nagy-Soper subtraction has
required several modifications, however, the technical description is
beyond the scope of this paper.
\begin{widetext}
\begin{equation}
\label{eq_NLO}
\begin{split}
    d\sigma^{\textrm{NLO}}_{t\bar{t}jj}&=\Gamma_{t}^{-2}
    \bigl(
    \overbrace{
    \left(d\sigma_{t\bar{t}jj}^{\textrm{LO}}+ d\sigma_{t\bar{t}jj}^{\textrm{virt}}+
    d\sigma_{t\bar{t}jjj}^{\textrm{real}}\right)
    \,d\Gamma_{t\bar{t}}^{\textrm{LO}}}^{{\rm Prod.}}
     +\overbrace{d\sigma_{t\bar{t}}^{\textrm{LO}}
    \,\left(d\Gamma_{t\bar{t}jj}^{\textrm{LO}}
    +d\Gamma_{t\bar{t}jj}^{\textrm{virt}}+
    d\Gamma_{t\bar{t}jjj}^{\textrm{real}}\right)}^{{\rm Decay}}
    \\&
     +\underbrace{ 
     d\sigma_{t\bar{t}j}^{\textrm{LO}}
    \,d\Gamma_{t\bar{t}j}^{\textrm{LO}}
     +
     d\sigma_{t\bar{t}jj}^{\textrm{LO}}
    \,d\Gamma_{t\bar{t}}^{\textrm{virt}}+
    d\sigma_{t\bar{t}}^{\textrm{virt}}\,
    d\Gamma_{t\bar{t}jj}^{\textrm{LO}}+
    d\sigma_{t\bar{t}j}^{\textrm{virt}}\,
    d\Gamma_{t\bar{t}j}^{\textrm{LO}}+
    d\sigma_{t\bar{t}j}^{\textrm{LO}}\,
    d\Gamma_{t\bar{t}j}^{\textrm{virt}}+
     d\sigma_{t\bar{t}jj}^{\textrm{real}}\,
    d\Gamma_{t\bar{t}j}^{\textrm{real}}+
    d\sigma_{t\bar{t}j}^{\textrm{real}}\,
    d\Gamma_{t\bar{t}jj}^{\textrm{real}}}_{{\rm Mix}}
    \bigl)
    \end{split}
\end{equation}
\end{widetext}
The set of subtraction terms needed in both schemes for 
calculations in the NWA is now completed and \textsc{Helac-Dipoles} 
can perform NLO QCD calculations for $pp \to t\bar{t}$ processes with an
arbitrary number of colourless/colourful final states.

We present results for the LHC with $\sqrt{s}=13$ TeV. The SM
parameters are given within the $G_\mu$-scheme with $G_{\mu} =1.166
378 7 \cdot 10^{-5} ~\textrm{GeV}^{-2}$, $m_Z = 91.1876$ GeV, $m_W
=80.379$ GeV and $\Gamma_W^{\textrm{NLO}} = 2.0972$ GeV.  The top
quark mass is set to $m_t = 172.5$ GeV.  All other particles,
including bottom quarks and leptons, are considered massless.  For the
LO and NLO top-quark width, based on
Refs. \cite{Jezabek:1988iv,Denner:2012yc}, we use the following values
\begin{equation}
\begin{split}
\Gamma^{\textrm{LO}}_{t} &=
1.4806842 ~{\rm GeV}\,, \\[0.2cm]
\Gamma^{\textrm{NLO}}_{t} & = 1.3535983 
~{\rm GeV} \,.
\end{split}
\end{equation}
We treat $\Gamma_t$ as a fixed parameter and its NLO value corresponds
to $\mu_R = m_t$ \footnote{While calculating the scale dependence for
the NLO cross section we keep $\Gamma_t^{\rm NLO}$ fixed. The error
introduced by this treatment is however small, and particularly for
two scales $\mu_R = 2  m_t$ and $\mu_R = m_t/2$ is at the level of 
$1\%$ only.}. We work in the 5-flavour scheme but neglect bottom
quarks in the initial state. We use an exclusive setup by requiring
exactly two $b$-jets in the final state, where we define a $b$-jet as
a jet with nonzero net bottomness. For example, a jet containing a
single $b$ quark (or $b$ anti-quark) plus a gluon is considered to be
a $b$ jet. On the contrary, a jet containing $b\bar{b}$ is labelled as
a light jet, see e.g. Ref.
\cite{Czakon:2020qbd,Bevilacqua:2021cit}. Therefore, in the real
emission part of the calculation we do not encounter sub-processes
with one bottom quark in the initial state like for example $bg\to
t\bar{t}bgg$ or $bg\to t\bar{t}bq\bar{q}$ where $q=u,d,c,s$ that would
lead to the underlining $bg\to t\bar{t}bg$ configuration at the born
level. On the other hand, $gg/q\bar{q} \to t\bar{t} b\bar{b}g$
sub-processes, with $gg/q\bar{q} \to t\bar{t} gg$ underlining born
contributions, are consistently included in our calculations.  The
remaining $b$-initiated sub-processes, $b\bar{b}/\bar{b}b\to
t\bar{t}gg$ and $b\bar{b}/\bar{b}b\to t\bar{t}q\bar{q}$, are about
$0.01\%$ of the total LO cross section and thus numerically
negligible. In order to check the correctness of our calculations, as
mentioned earlier, two independent subtraction schemes have been
employed. We found agreement on a per mille level between the results
obtained with the two schemes. The LHAPDF interface
\cite{Buckley:2014ana} is used to provide an access to PDFs and we
employ the NLO NNPDF3.1 PDF set \cite{NNPDF:2017mvq} at LO and
NLO. The running of the strong coupling constant is performed in both
cases with two-loop accuracy. We provide additionally the results for
MSHT20 \cite{Bailey:2020ooq} and CT18 \cite{Hou:2019efy}.  The three
PDF sets, which we employ, are recommended for SM calculations by the
PDF4LHC Working Group \cite{PDF4LHCWorkingGroup:2022cjn}.  To cluster
final state partons into jets we use the $anti$-$k_T$ jet algorithm
\cite{Cacciari:2008gp} with $R = 0.4$.  We require exactly two
opposite-sign charged leptons, two $b$-jets and at least two light
jets.  Leptons are required to have $p_{T \,\ell} > 20$ GeV, $|y_\ell|
< 2.4$, $\Delta R_{\ell\ell} > 0.4$ and $M_{\ell \ell} > 20$ GeV,
where $\ell^\pm = e^\pm, \mu^\pm$ as mentioned before.  Flavoured jets
with $p_{T \,b} > 30$ GeV and $|y_{b}| < 2.4$ are selected and only
$b$-jets that are well separated from leptons, $\Delta R_{b \ell}
>0.4$, are taken into account. Light jets are required to have $p_{T
\, j} > 40$ GeV and $|y_j| < 2.4$. They have to be isolated from
leptons and $b$-jets according to $\Delta R_{j \ell}> 0.4$ and $\Delta
R_{j b} > 0.8$.  The latter cut minimises gluon radiation from
top-quark decays.  The default renormalisation $(\mu_R)$ and
factorisation $(\mu_F)$ scale settings are
\begin{equation}
\mu_R=\mu_F=\mu_0=\frac{H_T}{2}\, ,
\end{equation}
where $H_T$ is calculated according to
%
\begin{table}
\begin{ruledtabular}
  \begin{tabular}{lllcc}
     $i$     & $\sigma^{\rm LO}$ [fb]       &
     $\sigma^{\rm NLO}$ 
     [fb] &    $\sigma^{\rm LO}_i/\sigma^{\rm LO}_{\rm Full}$ 
     & $\sigma^{\rm NLO}_i/\sigma^{\rm NLO}_{\rm Full}$ \\
 &&&&\\
\hline
Full & $868.8(2)^{\, +60\%}_{\, -35\%}$ 
& $~1225(1)^{\, ~+1\%}_{\, -14\%}$ 
& $1.00$ & $~~1.00$ \\
Prod. & $ 843.2(2)^{\, +60\%}_{\, -35\%} $ & $ ~1462(1)^{\,+12\%}_{\,-19\%}$
& $0.97$ &  $~~1.19$ \\
Mix & $25.465(5)$ & $-236(1)$ & $0.029$ & $-0.19$\\
Decay & $0.2099(1)$ & $0.1840(8)$ & $0.0002$& $~~0.0002$
  \end{tabular}
\end{ruledtabular}
  \caption{\it \label{tab:1a}   Integrated fiducial cross section at
    LO and NLO for the $pp \to t\bar{t}jj$ process at the
    LHC with $\sqrt{s}=13$ TeV.   Results are given for the default
    cuts with $\Delta R_{jb} > 0.8$. The full result as well as  {\it
      Prod.}, {\it Decay} and {\it Mix} contributions are
    shown. Theoretical uncertainties from scale variations and MC
  integration errors (in parentheses) are also displayed.}
\end{table}
\begin{table}
\begin{ruledtabular}
  \begin{tabular}{lllcc}
     $i$     & $\sigma^{\rm LO}$ [fb]       &
     $\sigma^{\rm NLO}$ 
     [fb] &    $\sigma^{\rm LO}_i/\sigma^{\rm LO}_{\rm Full}$ 
     & $\sigma^{\rm NLO}_i/\sigma^{\rm NLO}_{\rm Full}$ \\
 &&&&\\
\hline
Full & $1074.5(3)^{\, +60\%}_{\, -35\%}$ 
& $~1460(1)^{\, ~+1\%}_{\, -13\%}$ 
& $1.00$ & $~~1.00$ \\
Prod. & $ 983.1(3)^{\, +60\%}_{\, -35\%} $ & $ ~1662(1)^{\,+11\%}_{\,-18\%}$
& $0.91$ &  $~~1.14$ \\
Mix & $89.42(3)$ & $-205(1)$ & $0.083$ & $-0.14$\\
Decay & $1.909(1)$ & $2.436(6)$ & $0.002$& $~~0.002$
  \end{tabular}
\end{ruledtabular}
  \caption{\it \label{tab:1b}  Integrated fiducial cross section at LO
  and NLO for the $pp \to t\bar{t}jj$ process at the LHC with
  $\sqrt{s}=13$ TeV.   Results are given for the default cuts but with
  $\Delta R_{jb} > 0.4$.  The full result as well as  {\it Prod.},
  {\it Decay} and {\it Mix} contributions are shown. Theoretical
  uncertainties from scale variations and MC  integration errors (in
  parentheses) are also displayed. }
\end{table}
%
\begin{equation}
        H_T= \sum_{i=1}^2 p_{T {\ell_i}} + \sum_{i=1}^2 
        p_{T {j_i}} + \sum_{i=1}^2 p_{T {b_i}}+ p_T^{miss}\,,
\end{equation}
and $p_T^{miss}$ is the missing transverse momentum constructed of two
neutrinos which are present in the $pp\to t\bar{t}jj$ process
$(\nu_\ell, \bar{\nu}_{\ell})$. In all cases,  we estimate scale
uncertainties by performing a $7$-point scale variation around the
central value of the scale $(\mu_0)$. Specifically, we vary  $\mu_R$
and $\mu_F$ independently in the following range
\begin{equation}
\begin{split}
  \left(
\mu_R,\mu_F\right)
    = & 
[(2\mu_0,2\mu_0),(2\mu_0, \mu_0),(\mu_0,2\mu_0), 
(\mu_0,\mu_0), \\[0.2cm]
&
(\mu_0/2,\mu_0), (\mu_0,\mu_0/2),(\mu_0/2,\mu_0/2)]
    \,,
 \end{split}
\end{equation}
and choose the minimum and maximum of the resulting cross sections.

The integrated LO and NLO fiducial cross sections for $pp\to
t\bar{t}jj$ are shown in Table \ref{tab:1a}. The LO cross section is
dominated by the $gg$ channel with the $\sigma^{\rm LO}_{gg}=
561.1(2)$ fb $(65\%)$ contribution. It is followed by the $gq$ channel
with $\sigma^{\rm LO}_{gq}=272.6(1)$ fb $(31\%)$, where $gq$ stands
for $gq, qg, g\bar{q}$ and $\bar{q}g$. The smallest contribution comes
from the $q\bar{q}$ channel with $\sigma^{\rm LO}_{q\bar{q}}=35.10(1)$
fb $(4\%)$, where $q\bar{q} \in (q\bar{q}, \bar{q}q, qq,
\bar{q}\bar{q}, qq^\prime, \bar{q}\bar{q}^\prime, q \bar{q}^\prime,
\bar{q}q^\prime)$. The ${\cal K}$-factor for the full NWA calculation,
defined as ${\cal K}=\sigma^{\rm NLO}_{\rm Full}/\sigma^{\rm LO}_{\rm
Full}$, is ${\cal K}=1.41$. Thus, $\order{\alpha_s}$ corrections are
of medium size.  When NLO QCD corrections are included, a reduction in
the dependence on the unphysical $\mu_R$ and $\mu_F$ scales from
$60\%$ down to $14\%$, thus of more than a factor of $4$, is
observed. In order to examine how light jets are distributed we
additionally provide in Table \ref{tab:1a} {\it Prod.}, {\it Decay}
and {\it Mix} contributions defined according to Eq. \eqref{eq_LO} and
Eq. \eqref{eq_NLO}. At LO the dominant input comes from {\it Prod.}
$(97\%)$ and it is followed by {\it Mix} $(3\%)$, whereas the {\it
Decay} part is negligible.  The {\it Mix} contribution increases from
$3\%$ to $8\%$ when the $\Delta R_{jb}$ cut is set to $0.4$ instead of
$0.8$ (see Table \ref{tab:1b}). Nevertheless, this contribution might
be safely disregarded given that scale uncertainties at this
perturbative order are much larger. Once NLO QCD corrections are
included not only the relative size of {\it Mix} increases up to
$19\%$, but also its sign is reversed. This behaviour is driven by
radiative corrections to the decays of $t\bar{t}jj$, denoted by
$d\sigma^{\rm LO}_{t\bar{t}jj} \, d\Gamma_{t\bar{t}}^{\rm virt}$ and
 $d\sigma_{t\bar{t}jj}^{\rm real} \, d\Gamma^{\rm real
}_{t\bar{t}j}$ in Eq. \eqref{eq_NLO}, which are known to be negative
and increase (in absolute values) when the $\Delta R_{jb}$ cut is
reduced to $0.4$.  However, other components of the Mix part, notable
$d\sigma^{\rm LO}_{t\bar{t}j}\, d\Gamma^{\rm LO}_{t\bar{t}j}$ and its
direct NLO QCD corrections are positive and more sensitive to this
cut. Consequently, the overall size of {\it Mix} reduces to $14\%$ for
$\Delta R_{jb} > 0.4$ (see Table \ref{tab:1b}).

From the considerations above, we can conclude that neither the size
nor the sign of the {\it Mix} contribution can be reliably estimated
on the basis of LO predictions.  In addition, its sensitivity to
higher-order effects is very dependent on the fiducial phase-space
cuts. Finally, omitting the {\it Mix} part can lead to rather
misleading conclusions about the size of higher-order corrections and
theoretical uncertainties.  Thus, unless full NLO QCD corrections are
carefully incorporated into the NWA computations of $pp \to
t\bar{t}jj$, it is not clear to what extent various predictions
available in literature can be trusted. If only the {\it Prod.}
contribution to $t\bar{t}jj$ is taken into account, under question
could be not only the modelling of top-quark decays but also the
extrapolation of the $t\bar{t}jj$ fiducial predictions to the full
phase space.

To investigate this further, we have compared our NLO QCD result in
the full NWA $(\sigma^{\rm NLO}_{\rm Full})$ to the prediction that
includes NLO QCD corrections to $t\bar{t}jj$ production only
(i.e. with two light jets present in the production stage) while
top-quark decays are modelled with LO accuracy. We refer to this
result as $\sigma^{\rm NLO}_{\rm Prod.\, LO_{Decay}}$. For the larger
value of the $\Delta R_{jb}$ cut the two results are almost identical,
whereas for $\Delta R_{jb} > 0.4$ a difference of only $5\%$ is
observed.  We therefore conclude that when comparing to $\sigma^{\rm
NLO}_{\rm Prod.\, LO_{Decay}}$ the inclusion of the {\it Mix}
contribution at NLO leads to rather small differences in the central
value for our setup.  However, we gain in theoretical accuracy,
reducing scale uncertainties by about $5\%$.
%
\begin{figure}
  \begin{center}
   \includegraphics[width=0.45\textwidth]{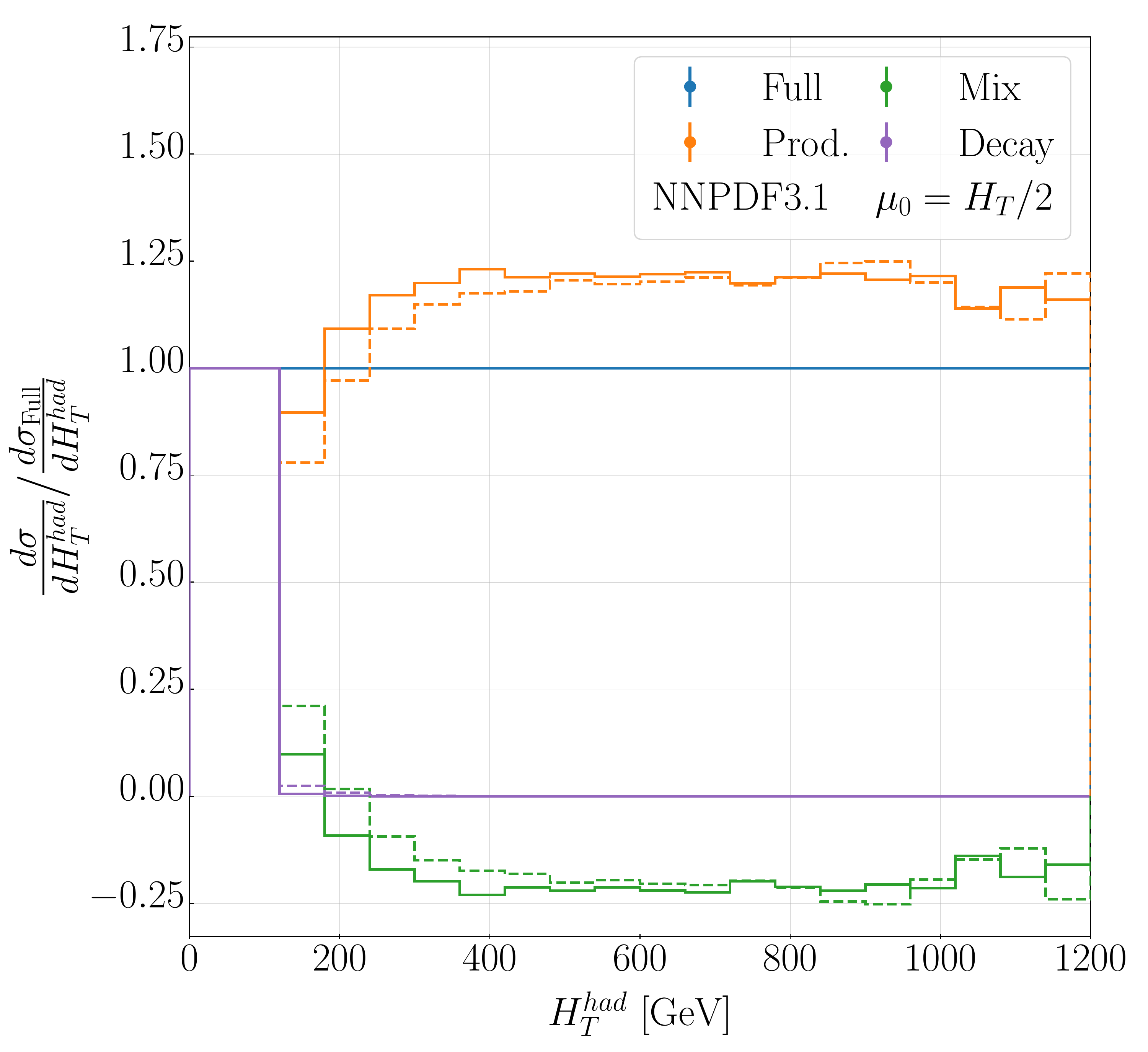}
   \includegraphics[width=0.45\textwidth]{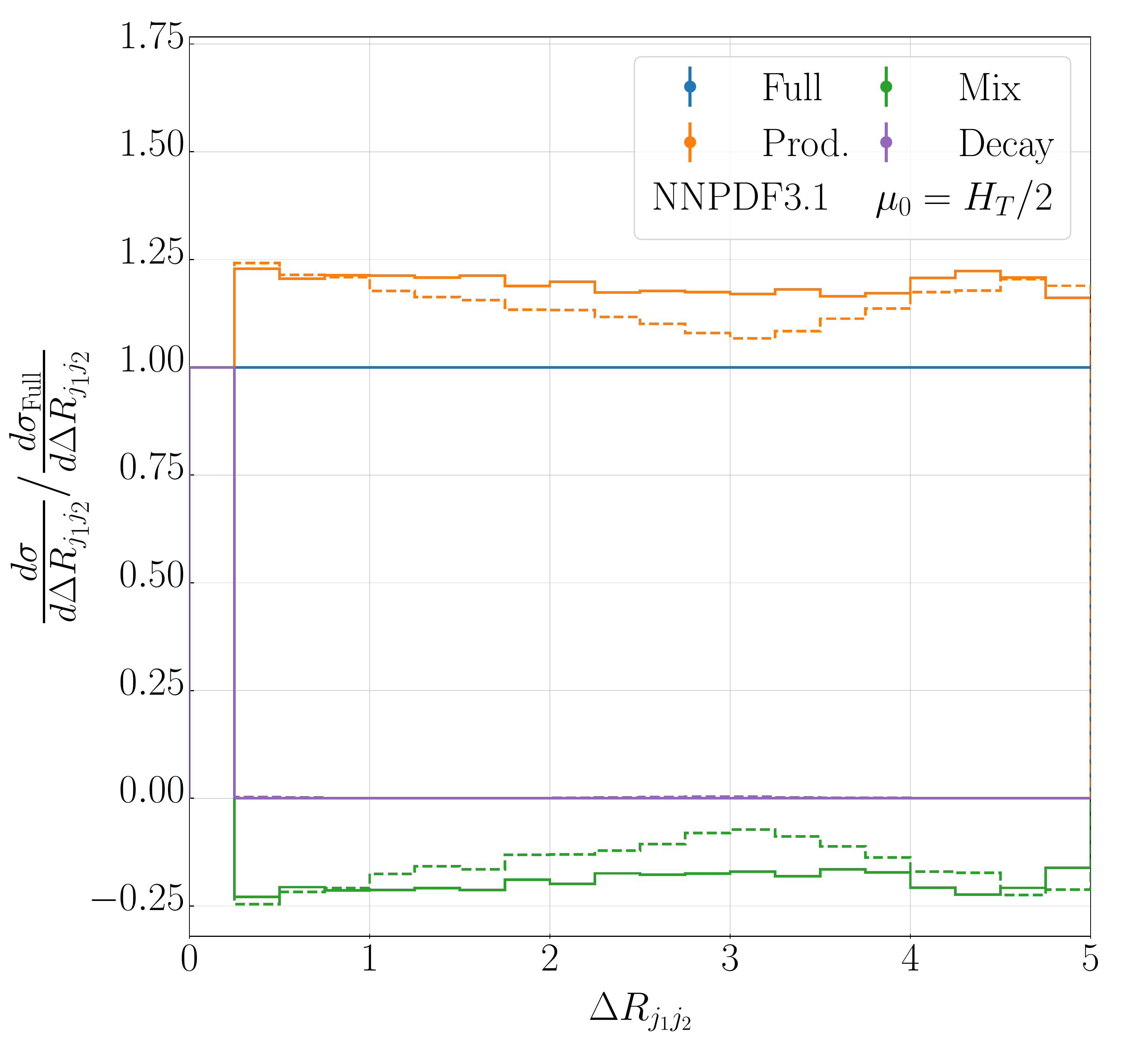}
\end{center}
\caption{\label{fig:fraction} 
\it Fraction of events when the two light jets come from $t\bar{t}$
production or  $t\to Wb$ decays as a function of $H_T^{had}$ and
$\Delta R_{j_1j_2}$. Also  shown is the mixed contribution where they
are emitted simultaneously in the production and top-quark decays.
NLO QCD results are shown for the $pp \to t\bar{t}jj$ process at the
LHC with $\sqrt{s}=13$ TeV. They are given for the default cuts with
$\Delta R_{jb} >0.8$ (solid line) and  $\Delta R_{jb} >0.4$ (dashed line).}
 \end{figure}

\begin{figure*}
  \begin{center}
  \includegraphics[width=0.45\textwidth]{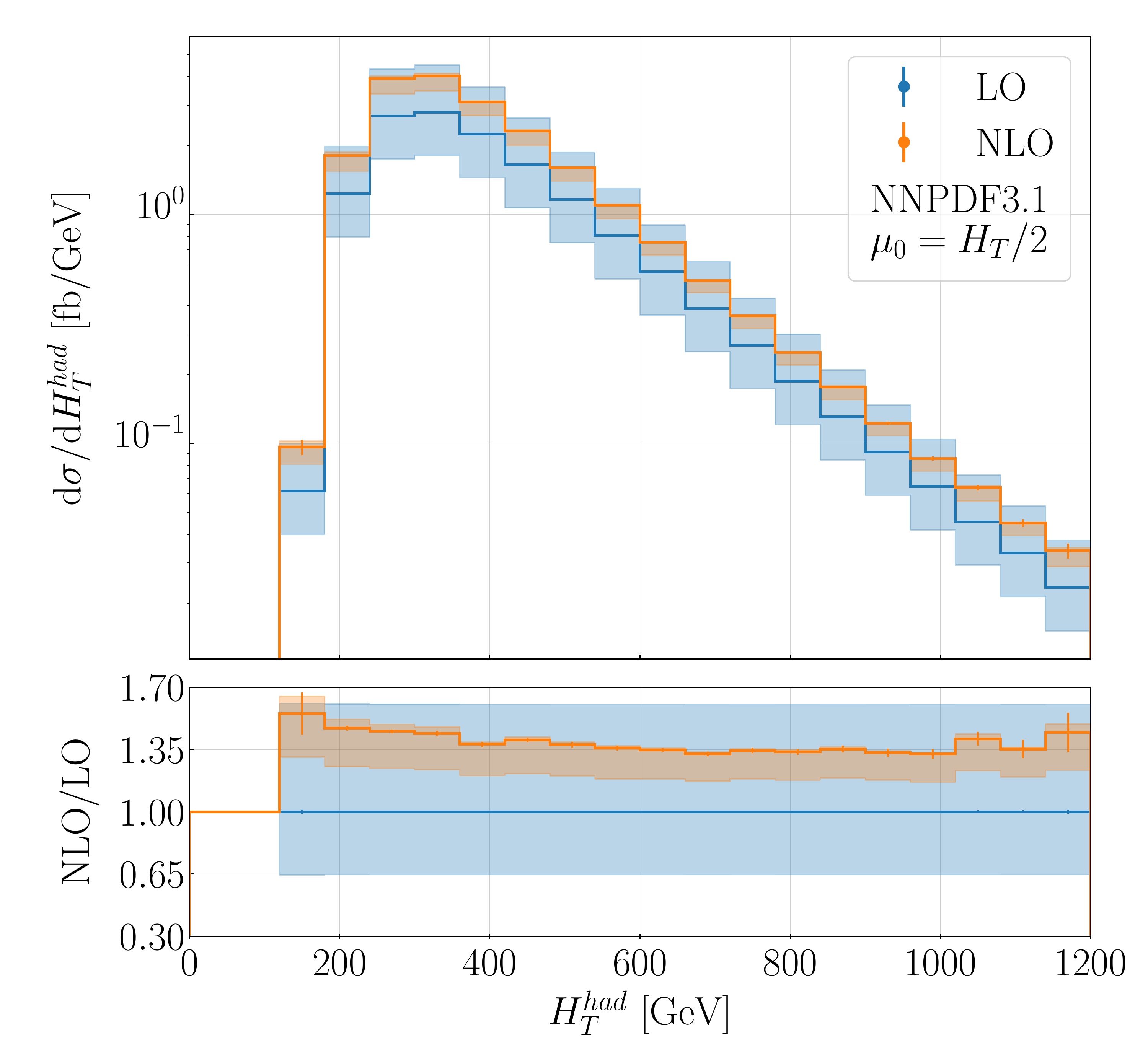}
    \includegraphics[width=0.45\textwidth]{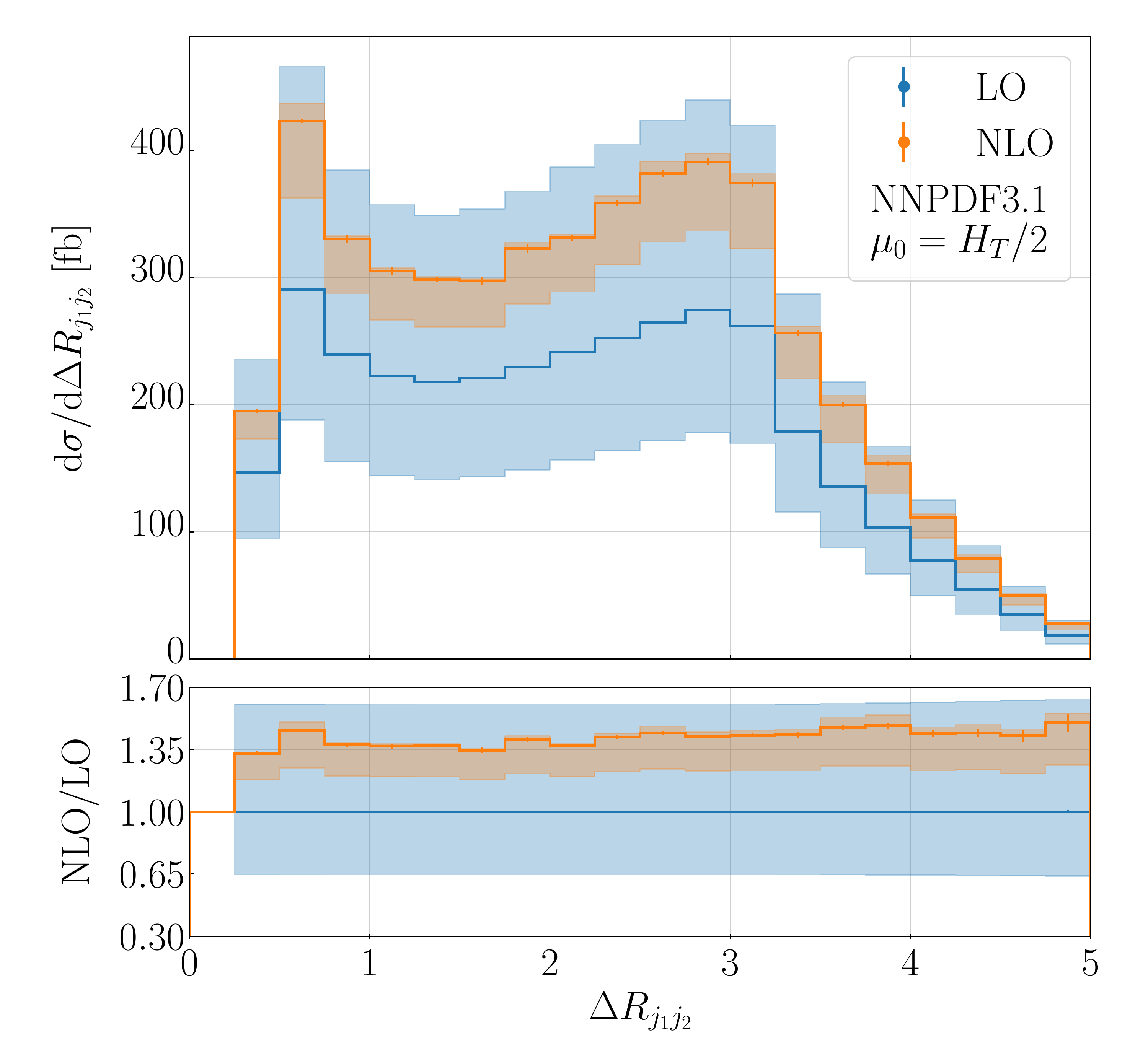}
\includegraphics[width=0.45\textwidth]{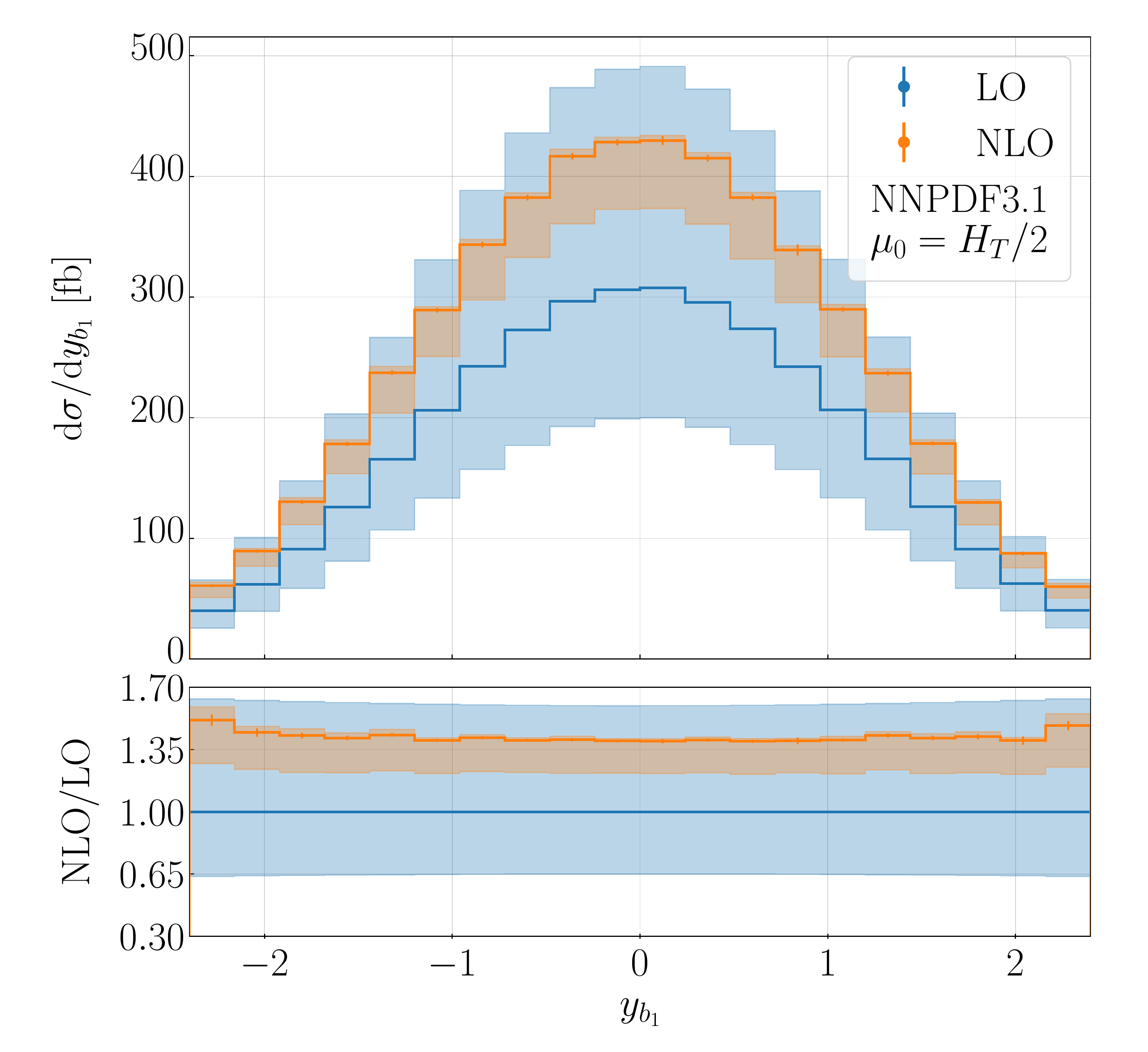}
    \includegraphics[width=0.45\textwidth]{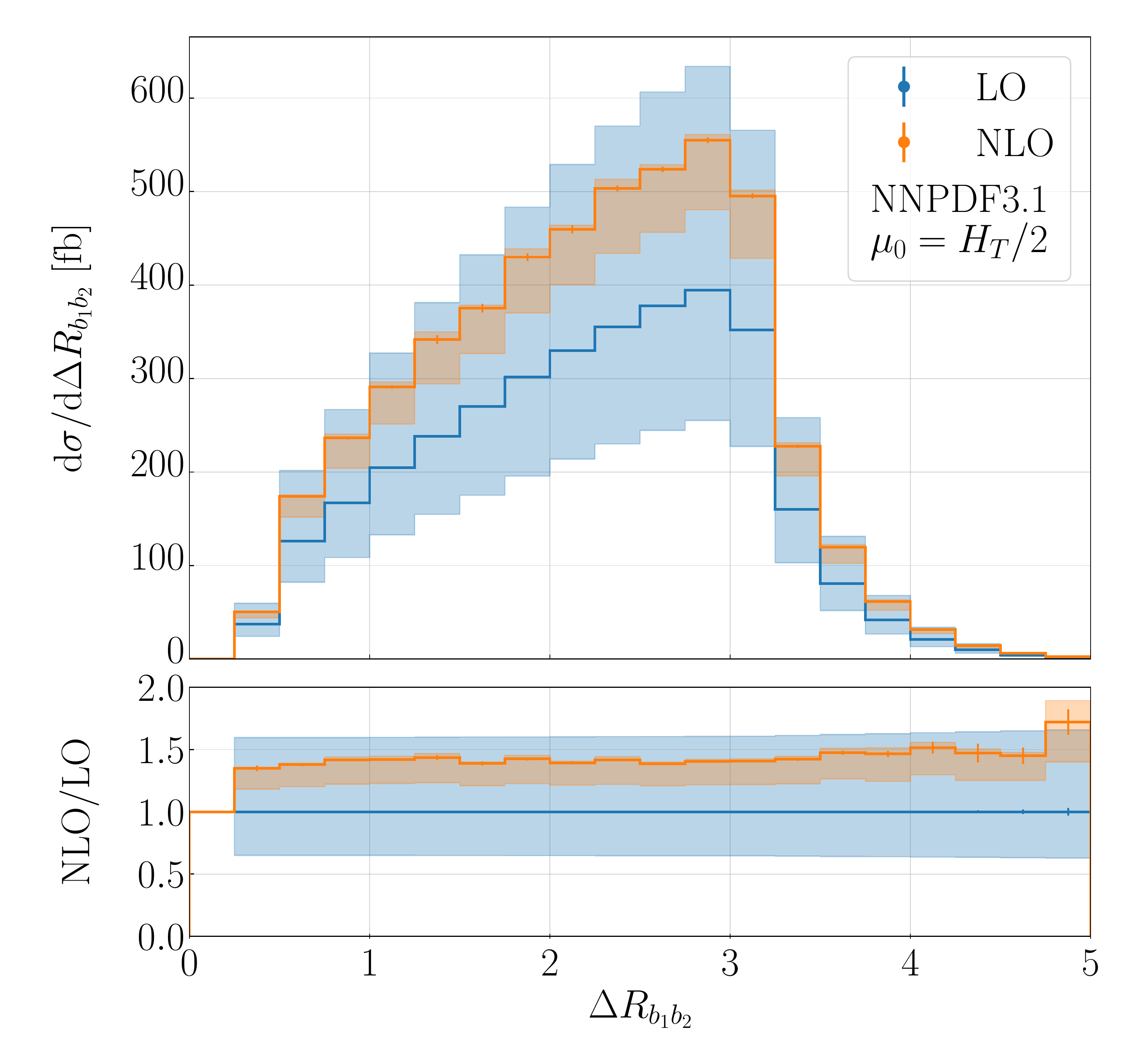}
\end{center}
\caption{\label{fig:diff} 
\it  Differential cross-section distributions at NLO in QCD as a
function of $H_T^{had}$,  $\Delta R_{j_1j_2}$, $y_{b_1}$ and $\Delta
R_{b_1b_2}$ for the $pp\to t\bar{t}jj$ process at the LHC with
$\sqrt{s}=13$ TeV. Results are given for the default cuts with $\Delta
R_{jb} > 0.8$.  The blue curve corresponds to the LO and the red curve
to the NLO result.  The corresponding uncertainty bands are also
shown. The lower panels display the differential ${\cal K}$-factor
together with uncertainty bands. }
 \end{figure*}

In Figure \ref{fig:fraction} we show the size of {\it Prod.}, {\it
Decay} and {\it Mix} relative to $\sigma^{\rm NLO}_{\rm Full}$ as a
function of $H_T^{had}$, defined as 
\begin{equation}
        H_T^{had} =  \sum_{i=1}^2 
        p_{T {j_i}} + \sum_{i=1}^2 p_{T {b_i}}\,,
\end{equation}
and $\Delta R_{j_1j_2}$. Also at the differential level the {\it
Decay} contribution is negligible. However, the negative {\it Mix}
contribution has nontrivial dependence on kinematics. For $H_T^{had}$
its importance increases (in absolute values) at the beginning of the
spectrum and from around $400$ GeV it stabilises at around $25\%$.
For $\Delta R_{j_1j_2}$ the {\it Mix} part is particularly important
for the $\Delta R_{j_1j_2} \le 1$ phase-space region, which represent
the bulk of the distribution (see also Figure \ref{fig:diff}),
especially when the default cut $\Delta R_{jb} > 0.8$ is
employed. Indeed, for small values of $\Delta R_{j_1j_2}$ the {\it
Mix} contribution amounts to $-23\%$, whereas for $\Delta R_{j_1j_2}
\approx 3$ we obtained $-17\%$.  Beyond that, the difference between
the two cases, $\Delta R_{jb} >0.8$ (solid line) and $\Delta R_{jb}
>0.4$ (dashed line), is clearly visible for the small $p_T$ region of
$H_T^{had}$ and for the back-to-back configurations in the case of
$\Delta R_{j_1j_2}$.  In the latter case, even though the {\it Mix}
contribution is still negative, it is reduced to $8\%$ in the case of
$\Delta R_{jb} > 0.4$.
%
\begin{figure}
  \begin{center}
     \includegraphics[width=0.45\textwidth]{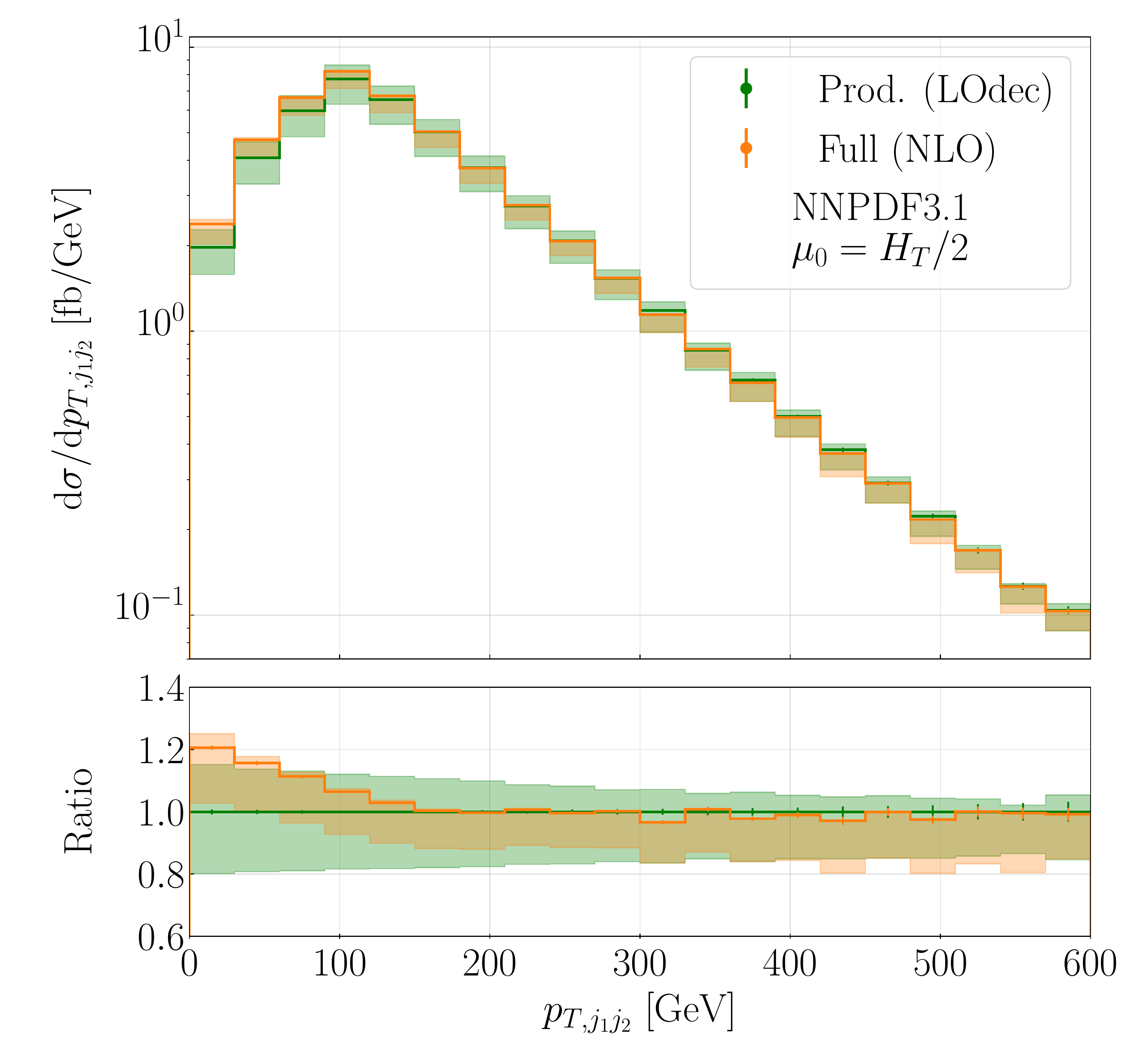}
      \includegraphics[width=0.45\textwidth]{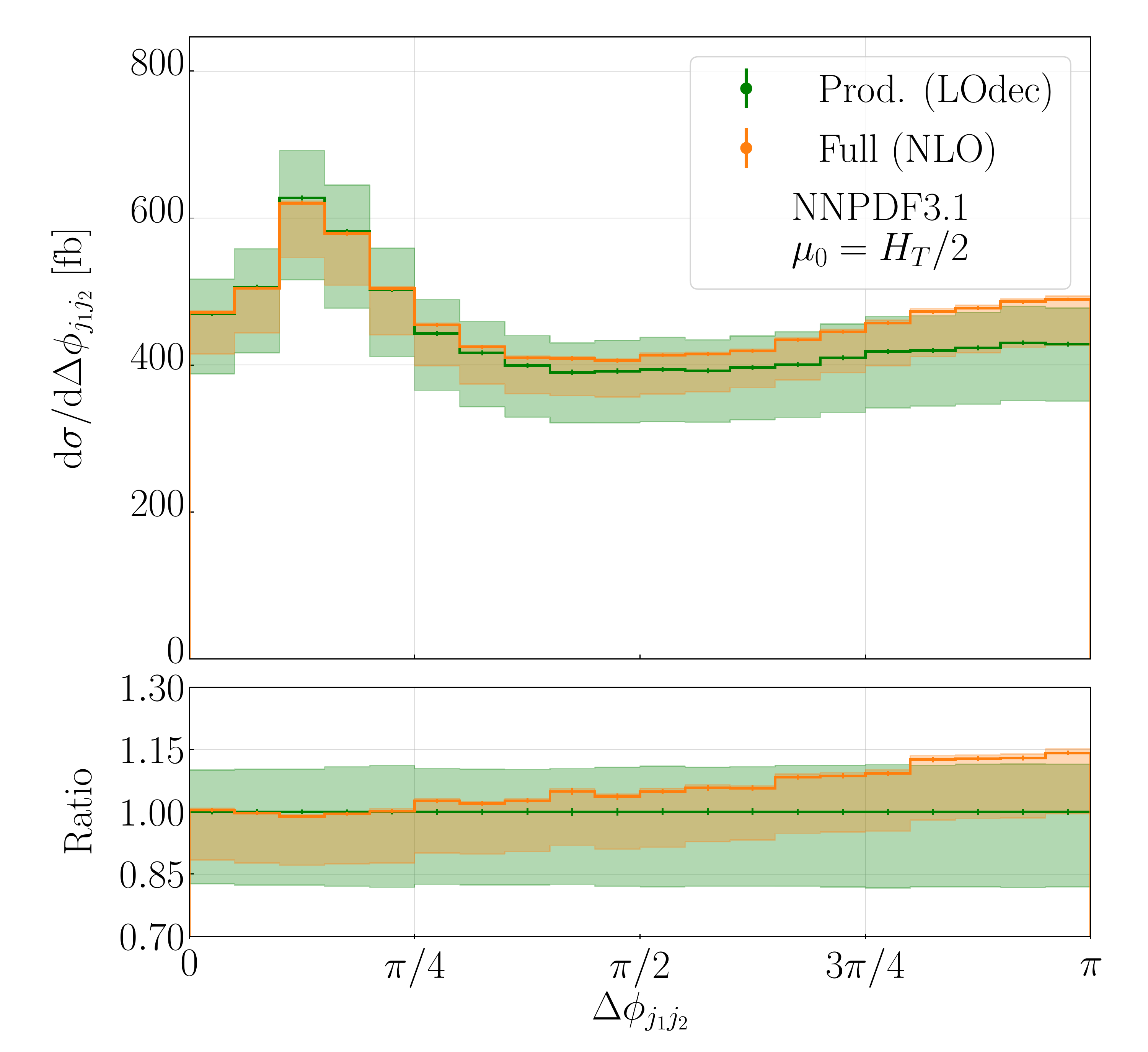}
\end{center}
\caption{\label{fig:LOdec} 
\it Differential cross-section distribution at NLO in QCD as a
function of $p_{T, \, j_1j_2}$ and $\Delta \phi_{j_1j_2}$ for the
$pp\to t\bar{t}jj$ process at the LHC with $\sqrt{s}=13$ TeV.  Results
are given for the default cuts but with $\Delta R_{jb} > 0.4$. Two
different theoretical descriptions are employed. The orange curve
corresponds to  the full NWA result and  the green curve to the NWA
prediction with LO top-quark decays  and two light jets in the
production stage only. The corresponding uncertainty bands are also
shown. The lower panels display the differential
$\sigma^{\rm NLO}_{\rm Full}/\sigma^{\rm NLO}_{\rm Prod.\,
  LO_{Decay}}$ ratio  with the corresponding uncertainty bands.  }
 \end{figure}
 
Also at the differential cross-section level, a comparison between
$\sigma^{\rm NLO}_{\rm Full}$ and $\sigma^{\rm NLO}_{\rm Prod.\,
LO_{Decay}}$ has been carried out. In this case, differences of up to
$15\%-20\%$ are observed for various observables that we have
examined. Both dimensionless and dimensionful observables as well as
various phase-space regions are affected.  As an example, we show in
Figure \ref{fig:LOdec} the transverse momentum of the $j_1j_2$ system
$(p_{T, \, j_1j_2})$ as well as the azimuthal angle difference between
these two light jets $(\Delta \phi_{j_1 j_2})$ at NLO in QCD for the
default cuts but with $\Delta R_{jb} > 0.4$. In the upper panels we
show the absolute NLO predictions for the full NWA and for the NWA
case with LO top-quark decays and two light jets in the production
stage only. In the lower panels the differential $\sigma^{\rm
NLO}_{\rm Full}/\sigma^{\rm NLO}_{\rm Prod.\, LO_{Decay}}$ ratio is
displayed together with the corresponding uncertainty bands.  The
error band is built bin-by-bin by employing, similarly to the
integrated fiducial cross-section case, a $7$-point scale
variation. In addition to the shape differences already mentioned, we
can observe reduced theoretical uncertainties for the full NWA result.
We can thus clearly see the importance of including the full NLO QCD
corrections in the NWA calculations for the $pp \to t\bar{t}jj$
process.

The second source of theoretical systematic uncertainties comprises
PDF uncertainties. For $pp\to t\bar{t} jj$ they are of the order of
$1.3\%$ for our default NNPDF3.1 PDF set.  Furthermore, for the MSHT20
PDF set we obtain $\sigma^{\rm NLO}_{\rm MSHT20} = 1212(1)$ fb,
whereas for CT18 we have $\sigma^{\rm NLO}_{\rm CT18} =1197(1)$
fb. The corresponding internal PDF uncertainties are $2.1\%$ and
$2.9\%$.  By comparing $\sigma^{\rm NLO}_{\rm NNPDF3.1}$ with
$\sigma^{\rm NLO}_{\rm MSHT20}$ and $\sigma^{\rm NLO}_{\rm CT18}$,
relative differences in the range of $1.1\%-2.3\%$ are observed, which
are consistent with the size of the internal PDF uncertainties.  In
general, PDF uncertainties are well below theoretical uncertainties
from scale variations.

While the size of the NLO QCD corrections to the integrated fiducial
cross section is certainly of interest, it is crucial to examine the
higher-order effects to various  differential
cross-section distributions.  In Figure \ref{fig:diff} we display four
examples: $H_T^{had}$, $\Delta R_{j_1j_2}$, $y_{b_1}$ (the hardest
$b$-jet's rapidity) and $\Delta R_{b_1 b_2}$.  Results are given for
the default cuts with $\Delta R_{jb} > 0.8$. For each plot the upper
panels show absolute LO and NLO predictions together with the
corresponding scale uncertainty bands. The lower panels display the
differential ${\cal K}$-factor. The LO and NLO uncertainty bands
normalised to the LO central value are also displayed. Similar to the
integrated fiducial cross sections we find a significantly reduced
dependence on the choice of $\mu_F$ and $\mu_R$ also at the
differential level. Specifically, for all observables we obtain LO
theoretical uncertainties of the order of $60\%$ while at NLO they are
maximally up to $15\%$, giving a reduction by a factor of $4$.
Moreover, scale dependence bands for LO and NLO predictions overlap
nicely, indicating a well behaved perturbative convergence.  It should
be noted here that we observe rather asymmetric theoretical
uncertainties analogous to the case of the integrated fiducial cross
section.  In such a situation, the maximum of the two values in each
bin should be taken as the final theoretical error estimate. An
alternative would be to symmetrise the two values, but this approach
might underestimate the final error, which will only be known once
NNLO QCD corrections to the $pp \to t\bar{t}jj$ process, with all
three contributions {\it Prod.}, {\it Decay} and {\it Mix} included,
are available.  We would like to emphasise here that the scale
variation is by no means a rigorous way to assess the true theoretical
uncertainty. At best, it might only give an indication of the full
uncertainty which is due to the not yet calculated higher order
corrections. As for the NLO QCD corrections to the differential
cross-section distributions, they are significant as they are in the
range of $30\% -50\%$.  An appropriate global ${\cal K}$-factor cannot
therefore be applied to all LO predictions to well approximate NLO
predictions. Consequently, the complete NLO QCD corrections should be
consistently incorporated to all differential cross-section
distributions.
%
\begin{table}
\begin{ruledtabular}
  \begin{tabular}{llll}
     ${\cal R}_n$     & ${\cal R}^{\rm LO}$     &
     ${\cal R}^{\rm NLO}$ 
     &   ${\cal R}^{\rm NLO}_{\rm exp}$   \\
 &&&\\
\hline
${\cal R}_1=\sigma_{t\bar{t}j}/\sigma_{t\bar{t}}$ 
& $0.3686 {}^{\,+12\%}_{\,-10\%} $ 
&  $0.3546 {}^{\, +0\%}_{\, -5\%} $
& $0.3522 {}^{\, +0\%}_{\,-3\%} $ \\
${\cal R}_2=\sigma_{t\bar{t}jj}/\sigma_{t\bar{t}j}$
& $0.2539 {}^{\, +11\%}_{\, ~-9\%}$
& $0.2660 {}^{\, +0\%}_{\, -5\%}$ 
& $0.2675 {}^{\, +0\%}_{\, -2\%}$ 
\\
  \end{tabular}
\end{ruledtabular}
 \caption{\it \label{tab:2}  LO and NLO cross section ratios for the
   $pp \to t\bar{t}jj$ process at the LHC with $\sqrt{s}=13$
   TeV. Results are given for the default cuts with $\Delta R_{jb} >
   0.8$. In the last column the expanded NLO cross section ratio to
   first order in  $\alpha_s$ is also given.  Theoretical
   uncertainties from scale variations, which are taken as correlated,
   are also displayed. MC integration errors are at the per mill level.}
\end{table}

To show the potential of \textsc{Helac-Nlo} we present results for
$pp\to t\bar{t}+ nj$ in the di-lepton channel, with $n=0,1,2$, in the
form of fiducial cross section ratios, defined as ${\cal R}_n
=\sigma_{t\bar{t}+nj}/\sigma_{t\bar{t}+(n-1)j}$.  They are displayed
in Table \ref{tab:2} for our default setup up to a modification in the
definition of $\mu_0=H_T/2$ for $pp\to t\bar{t}(j)$.  Theoretical
uncertainties from scale variations are taken as correlated. The
internal PDF uncertainties are evaluated in a similar fashion as for
the integrated fiducial cross section that allow us to properly
account for cross-correlations between the two processes considered
for the numerator and the denominator of ${\cal
R}_{1,2}$. Nevertheless, they are only up to $0.5\%$. The last column,
${\cal R}^{\rm NLO}_{\rm exp}$, shows a consistent expansion of ${\cal
R}$ in $\alpha_s$. In both cases, the difference between ${\cal
R}^{\rm NLO}_{\rm exp}$ and ${\cal R}^{\rm NLO}$ is similar in size as
the MC errors. Thus, ${\cal R}_{1,2}$ are very stable and precise
observables that should be measured at the LHC. Indeed, a judicious
choice of $\mu_R, \mu_F$ allows us to obtain $2\% - 3\%$ precision for
${\cal R}^{\rm NLO}_{1,2}$. Until now, such precise predictions were
reserved only for $pp\to t\bar{t}$ at NNLO QCD.

\section*{Conclusions}

We have computed, for the first time, the complete set of NLO QCD
corrections to $pp\to t\bar{t}jj$ including top-quark decays at the
LHC in the NWA. Our calculation shows that NLO QCD corrections to
$pp\to t\bar{t}jj$ with realistic final states play an important role
as they substantially increase the SM prediction and significantly
decrease the dominant scale uncertainty. An important finding of this
paper is the magnitude and the sign of the {\it Mix} contribution. The
latter is normally omitted in various studies for $pp\to
t\bar{t}jj$. As important as it is, if not properly accounted for, it
can affect the modelling of top-quark decays and the extrapolation to
full phase space with existing MC event generators. This is especially
true when considering various differential cross-section distributions
where differences between $\sigma^{\rm NLO}_{\rm Full}$ and
$\sigma^{\rm NLO}_{\rm Prod. \,LO_{Decay}}$ up to $15\% - 20\%$ are
observed.

We conclude by saying that it would be beneficial to make a comparison
between the results obtained in this work and those reported in
literature. Such a comparison could assess the extent to which parton
shower effects can reproduce all the contributions required at the NLO
level in QCD for the $pp \to t\bar{t}jj$ process. In addition, it
could help to identify regions of phase space for specific observables
that are indeed sensitive to resummed dominant soft-collinear
logarithmic corrections from parton showers, that are absent in our
fixed-order predictions for this process. We plan to carry out such
comparisons in the future.

\begin{acknowledgments}
The work of M.L., D.S. and M.W. was supported by the Deutsche
Forschungsgemeinschaft (DFG) under grant 396021762 - TRR 257: {\it P3H
- Particle Physics Phenomenology after the Higgs Discovery} and by the
DFG under grant 400140256 - GRK 2497: {\it The physics of the heaviest
particles at the Large Hardon Collider.}

Support by a grant of the Bundesministerium f\"ur Bildung und
Forschung (BMBF) is additionally acknowledged.

G.B. was supported by the Hellenic Foundation for Research
and Innovation (H.F.R.I.) under the “2nd Call for H.F.R.I. Research
Projects to support Faculty Members $\&$ Researchers” (Project Number:
02674 HOCTools-II).

Simulations were performed with computing resources granted by RWTH
Aachen University under projects {\tt p0020216},  {\tt rwth0414} and
{\tt rwth0846}.

\end{acknowledgments}


\end{document}